\documentclass[showpacs,twocolumn,preprintnumbers,amsmath,amssymb]{revtex4}

\usepackage{graphicx}
\usepackage{dcolumn}
\usepackage{bm}
\usepackage{float}
\usepackage{multirow}
\usepackage{hhline}

\newcommand{\bq}{\begin{equation}}
\newcommand{\eq}{\end{equation}}
\newcommand{\bqa}{\begin{eqnarray}}
\newcommand{\eqa}{\end{eqnarray}}
\newcommand{\nn}{\nonumber \\}

\def\be     {\begin{equation}}
\def\ee     {\end{equation}}
\def\bea        {\begin{eqnarray}}
\def\eea        {\end{eqnarray}}
\def\bnn    {\begin{eqnarray*}}
\def\enn    {\end{eqnarray*}}

\begin{document}

\title{Anomalous Hall effects beyond Berry magnetic fields in a Weyl metal phase}
\author{Iksu Jang, Jaeho Han, and Ki-Seok Kim}
\affiliation{Department of Physics, POSTECH, Pohang, Gyeongbuk 790-784, Korea}

\date{\today}

\begin{abstract} 
Applying time-varying magnetic fields to Weyl metals, a pair of Weyl points become oscillating. This oscillating monopole and anti-monopole pair gives rise to AC Berry magnetic fields, responsible for the emergence of Berry electric fields, which have not been discussed before at least in the context of Weyl metals. Introducing this novel information into Boltzmann transport theory, we find anomalous Hall effects beyond Berry magnetic fields as a fingerprint of Berry electric fields.
\end{abstract}

%\pacs{71.10.Hf, 71.30.+h, 71.10.-w, 71.10.Fd}

\maketitle

\section{Introduction}

Recently, Weyl metals \cite{WM1,WM2,WM3} are of interests not only in condensed matter physics but also in particle physics, involved with their topologically identified nontrivial properties \cite{WM_Reviews}. Their anomalous metallic properties originate from a pair of Weyl bands, separated in momentum space, where each Weyl band describes emergent relativistic Weyl electrons. The band structure itself may be regarded to be a three dimensional version of a graphene. Interestingly, the pair of Weyl points can be identified with a magnetic monopole and anti-monopole pair in momentum space. Accordingly, the Berry curvature, or more accurately, the Berry magnetic field is assigned by this monopole pair, which turns out to play an essential role in anomalous transport phenomena of Weyl metals \cite{NLMR_First_Exp,NLMR_Followup_I,NLMR_Followup_II,NLMR_Followup_III,Nielsen_Ninomiya_NLMR,CME1,CME2,CME3,CME4,CME5,CME6,CME7,Boltzmann_Chiral_Anomaly1,Boltzmann_Chiral_Anomaly2,
Boltzmann_Chiral_Anomaly3,Boltzmann_Chiral_Anomaly4,Boltzmann_Chiral_Anomaly5,Boltzmann_Chiral_Anomaly6,AHE1,AHE2,AHE3,Disordered_Weyl_Metal1,Disordered_Weyl_Metal2}.

The present work starts from an idea that the relative position of the monopole pair can be controlled by external magnetic fields \cite{WM_Reviews}. Applying time-varying magnetic fields to Weyl metals, we investigate the role of an oscillating monopole pair in the transport. This oscillating monopole pair is expected to cause AC Berry magnetic fields \cite{Shindou_Berry_Electric_Field}. If such Berry magnetic fields are governed by Berry-Maxwell equations, the Maxwell equation in momentum space, Berry electric fields would be generated. In this study we investigate the role of the emergent Berry electric field in the transport of Weyl metals beyond the Berry magnetic field.

Based on Boltzmann transport theory with a topologically modified Drude model, which takes into account the novel information of the Berry electric field, we find that the oscillating monopole pair gives rise to anomalous Hall currents. These anomalous Hall currents should be distinguished from ``conventional" anomalous Hall currents described by Berry magnetic fields. We classify these Hall currents in all possible situations. We reveal that these anomalous Hall effects are involved with an extended chiral anomaly given by a field theory, where a time-varying chiral gauge field appears to describe an ``oscillating" relative-distance vector of the monopole pair in the anomaly equation.

\section{Chiral gauge field as the gradient of an axion angle of the topological-in-origin $\bm{E} \cdot \bm{B}$ term}

The chiral anomaly \cite{Peskin_Schroeder} is an essential ingredient in anomalous transport phenomena of Weyl metals \cite{Nielsen_Ninomiya_NLMR}. It is encoded into an inhomogeneous topological-in-origin $\theta-$term \cite{Axion_EM}, given by
\bqa && Z = \int D \psi_{\alpha a} \exp\Big[- \int_{0}^{\beta} d \tau \int d^{3} \bm{r} \nn && \Big\{ \psi_{\alpha a}^{\dagger} \Big( (\partial_{\tau} - \mu) \bm{I}_{\alpha\beta} \otimes \bm{I}_{ab} - i v_{D} (\bm{\partial}_{\bm{r}} - i \bm{A}) \cdot \bm{\sigma}_{\alpha\beta} \otimes \bm{\tau}_{ab}^{z} \nn && + m \bm{I}_{\alpha\beta} \otimes \bm{\tau}_{ab}^{x} \Big) \psi_{\beta b} + \frac{\theta(\bm{r})}{2\pi} \frac{\alpha}{2\pi} \bm{E} \cdot \bm{B} \Big\} \Big] . \label{TFL_Dirac_Theory} \eqa
Here, $\psi_{\alpha a}$ is a four-component Dirac spinor with spin $\alpha$ and orbital $a$. $\bm{\sigma}_{\alpha\beta}$ and $\bm{\tau}_{ab}$ are two-by-two Pauli matrices, acting on spin and orbital spaces, respectively. $v_{D}$ is a velocity, $m$ is a mass parameter, and $\mu$ is a chemical potential. $\bm{A}$ is an electromagnetic vector potential, regarded to be externally applied.
$\bm{E} = - \frac{1}{c} \partial_{\tau} \bm{A}$ and $\bm{B} = \bm{\nabla} \times \bm{A}$ are externally applied electric field and magnetic field, respectively. $\theta(\bm{r})$ is an axion angle, and $\alpha$ is a fine structure constant.

One can represent this effective theory in terms of four-by-four Dirac gamma matrices, given by
\bqa && \gamma^{0} = \bm{I}_{\alpha\beta} \otimes \bm{\tau}_{ab}^{x} , ~~~~~ \gamma^{k} = - i \bm{\sigma}_{\alpha\beta}^{k} \otimes \bm{\tau}_{ab}^{y} . \eqa
Then, the partition function reads
\bqa && Z = \int D \psi \exp\Big[- \int_{0}^{\beta} d \tau \int d^{3} \bm{r} \Big\{ \bar{\psi} \Big( i \gamma^{0} (\partial_{\tau} - \mu) \nn && - i v_{D} \bm{\gamma} \cdot (\bm{\partial}_{\bm{r}} - i \bm{A}) + m \Big) \psi_{\beta b} + \frac{\theta(\bm{r})}{2\pi} \frac{\alpha}{2\pi} \bm{E} \cdot \bm{B} \Big\} \Big] \eqa
with $\bar{\psi} = \psi^{\dagger} \gamma^{0}$.

In order to determine the angle parameter, we recall the chiral anomaly equation \cite{Peskin_Schroeder}
\bqa && \partial_{\mu} (\bar{\psi} \gamma^{\mu} \gamma^{5} \psi) = \frac{\alpha}{4 \pi^{2}} \bm{E} \cdot \bm{B} . \eqa
This equation states that the classically conserved chiral current is not preserved any more in the quantum level, described by applied electromagnetic fields. Replacing the topological-in-origin $\bm{E} \cdot \bm{B}$ term with the chiral current based on this anomaly equation and performing the integration by parts, we rewrite the effective action as follows
\bqa && \mathcal{S}_{eff} = \int_{0}^{\beta} d \tau \int d^{3} \bm{r} \Big\{ \bar{\psi} \Big( i \gamma^{0} (\partial_{\tau} - \mu) \nn && - i v_{D} \bm{\gamma} \cdot (\bm{\partial}_{\bm{r}} - i \bm{A} - i \gamma^{5} \bm{c}) + m \Big) \psi_{\beta b} \Big\} . \label{Weyl_Metal_Matter_Sector} \eqa
Here, $\bm{c}$ is a chiral gauge field, given by
\bqa && \bm{c} = \bm{\nabla}_{\bm{r}} \theta(\bm{r}) , \label{Chiral_Gauge_Field} \eqa
where $\gamma^{5} = i \gamma^{0} \gamma^{1} \gamma^{2} \gamma^{3} = \bm{I}_{\alpha\beta} \otimes \bm{\tau}_{ab}^{z}$ is the chiral matrix.
%
%It is essential to observe that the chiral gauge-field term is nothing but the Zeeman term, given by
%\bqa && \mathcal{S}_{Z} = - \int_{0}^{\beta} d \tau \int d^{3} \bm{r} v_{D} g \psi_{\alpha a}^{\dagger} [\bm{B} \cdot (\bm{\sigma}_{\alpha\beta} \otimes \bm{I}_{ab})] \psi_{\beta b} , \nn \eqa
%if this term is expressed in terms of Dirac gamma matrices. The chiral gauge field is identified with the external magnetic field
%\bqa && \bm{c} = g \bm{B} , \eqa
%where $g$ is the Lande $g-$factor. As a result, the axion angle is
%\bqa && \theta(\bm{r}) = g \bm{B} \cdot (\bm{r} + \bm{R}) , \eqa
%where $\bm{R}$ describes the freedom of a reference point.
%
It turns out that the band structure of this effective action describes that of a Weyl metal phase, where the distance between a pair of Weyl points is given by $2 \bm{c}$ in the case of $m = 0$ \cite{WM_Reviews}.

\section{Maxwell equations in momentum space}

In order to describe the dynamics of electromagnetic fields in Weyl metals, we start from the following effective action for electromagnetic fields \cite{Axion_EM}
\bqa \mathcal{S}_{EM} &=& \int d t d^{3} \bm{r} \Big\{ \frac{1}{8 \pi} (\bm{E}^{2} - \bm{B}^{2}) + \frac{\theta}{2\pi} \frac{\alpha}{2\pi} \bm{E} \cdot \bm{B} \nn &-& \frac{1}{c} \bm{j} \cdot \bm{A} - \rho \Phi \Big\} . \eqa
Here, both electric and magnetic fields are given by
\bqa && \bm{E} = - \bm{\nabla} \Phi - \frac{1}{c} \partial_{t} \bm{A}, ~~~~~ \bm{B} = \bm{\nabla} \times \bm{A} , \eqa
respectively, where $\bm{A}$ and $\Phi$ are electromagnetic vector and scalar potentials. The topological-in-origin $\bm{E} \cdot \bm{B}$ term breaks time reversal symmetry generally speaking, encoding chiral anomaly. $\alpha$ is a fine structure constant, as introduced before. $\bm{j}$ and $\rho$ represent electrical current and charge density, respectively. Dynamics of these matter fields are described by Eq. (\ref{Weyl_Metal_Matter_Sector}).

Applying the variational principle to this effective action with respect to electromagnetic vector and scalar potentials, we obtain modified Maxwell equations to describe axion electrodynamics \cite{Axion_EM}
\bqa && \bm{\nabla} \cdot \bm{E} = 4 \pi \rho + \frac{\alpha}{\pi} (\bm{\nabla} \theta \cdot \bm{B}) , \nn && \bm{\nabla} \times \bm{B} - \frac{1}{c} \partial_{t} \bm{E} = \frac{4\pi}{c} \bm{j} - \frac{\alpha}{\pi} \Big( (\bm{\nabla} \theta \times \bm{E}) + \frac{1}{c} (\partial_{t} \theta) \bm{B} \Big) , \nn && \bm{\nabla} \times \bm{E} + \frac{1}{c} \partial_{t} \bm{B} = 0 , ~~~~~ \bm{\nabla} \cdot \bm{B} = 0 . \eqa
If we redefine both electric and magnetic fields as follows $\bm{\mathbb{E}} = \bm{E} - \frac{\alpha}{\pi} \theta \bm{B}$ and $\bm{\mathbb{B}} = \bm{B} + \frac{\alpha}{\pi} \theta \bm{E}$, respectively, these equations are reduced to conventional Maxwell equations to describe Maxwell electrodynamics. In other words, the topological-in-origin $\bm{E} \cdot \bm{B}$ term gives rise to mixing between electric and magnetic fields.

In order to understand the axion electrodynamics, we should find how both the electrical current and charge density are represented in terms of electric and magnetic fields, referred to as constituent equations. An essential point is that the ``conventional" Ohm's law does not work in Weyl metals \cite{Nielsen_Ninomiya_NLMR,CME1,CME2,CME3,CME4,CME5,CME6,CME7,Boltzmann_Chiral_Anomaly1,Boltzmann_Chiral_Anomaly2,Boltzmann_Chiral_Anomaly3,Boltzmann_Chiral_Anomaly4, Boltzmann_Chiral_Anomaly5,Boltzmann_Chiral_Anomaly6,AHE1,AHE2,AHE3,Disordered_Weyl_Metal1,Disordered_Weyl_Metal2}. Novel constituent equations should be uncovered. Actually, they can be found, based on Boltzmann transport theory for Weyl metals
\bqa && \partial_{t} f_{\chi} + \dot {\bm r}_\chi \cdot \bm{\nabla}_{\bm{r}} f_{\chi} + \dot {\bm p}_\chi \cdot \bm{\nabla}_{\bm{p}} f_{\chi} = - \frac{f_{\chi} - f_{eq}}{\tau_{eff}} . \label{Boltzmann_Equation} \eqa
Here, $f_{\chi} = f_{\chi}(\bm{p};\bm{r},t)$ is a distribution function of chiral fermions near a chiral Fermi surface, given by the chirality $\chi = \pm$, where $\bm{p}$ is a momentum, the Fourier transformed coordinate of a relative distance between a particle-hole pair, and $\bm{r}$ and $t$ are center-of-mass coordinates of the particle-hole pair \cite{Boltzmann_LFL}. $f_{eq} = f_{eq}(\bm{p})$ is an equilibrium distribution function. $\tau_{eff}$ is an effective relaxation time in terms of disorder scattering between intra Fermi surfaces of the same chirality and that between inter Fermi surfaces of the opposite chirality.

The effective velocity of $\dot {\bm r}_\chi$ and the effective force of $\dot {\bm p}_\chi$ are described by
\bqa && \dot {\bm r}_\chi = {\bm v}_{\chi} + \chi \dot{\bm c} \times \boldsymbol{\mathcal{B}}_{\chi} + \dot {\bm p}_\chi \times \boldsymbol{\mathcal{B}}_{\chi} \label{Effective_Velocity_Weyl_Metal} \eqa
and
\bqa && \dot {\bm p}_\chi = e {\bm E} + \frac{e}{c} \dot {\bm r}_\chi \times {\bm B} , \label{Effective_Force_Weyl_Metal} \eqa
respectively. If the second and third terms are neglected in Eq. (\ref{Effective_Velocity_Weyl_Metal}), these two equations are referred to as the Drude model. Here, ${\bm v}_{\chi}$ is the group velocity. The equation for $\dot {\bm p}_\chi$ describes the Lorentz force. On the other hand, the third term in Eq. (\ref{Effective_Velocity_Weyl_Metal}) gives rise to the contribution of anomalous velocity, where $\boldsymbol{\mathcal{B}}_{\chi}$ is Berry magnetic field \cite{Berry_Curvature_Review1,Berry_Curvature_Review2}. The second term is our main discovery, describing the Berry electric field. This contribution will be derived in the next section.

Resorting to this topologically modified Boltzmann transport theory, one can find constituent equations for charge density and electric current as follows
\bqa && \rho = \sum_{\chi} \rho_{\chi} \equiv e\sum_{\chi} \int \frac{d^{3} \bm{p}}{(2\pi)^{3}} G_{\chi} f_{\chi} , \nn && \bm{j} = \sum_{\chi} \bm{j}_{\chi} \equiv e \sum_{\chi} \int \frac{d^{3} \bm{p}}{(2\pi)^{3}} G_{\chi} \bm{\dot{r}}_{\chi} f_{\chi} , \label{Density_Current_Weyl_Metal} \eqa where the phase-space measure is modified as
\bqa && G_{\chi} = 1 + \frac{e}{c} \bm{B} \cdot \boldsymbol{\mathcal{B}}_{\chi} . \eqa
We recall $\sum_{\chi} \chi = 0$.

It is not surprising to observe that both Berry magnetic and Berry electric fields in the topologically modified Drude model satisfy Maxwell equations in momentum space, described by
\bqa && \bm{\nabla}_{\bm{p}} \cdot \boldsymbol{\mathcal{B}} = 2\pi\sum_{\chi} \chi \delta\Big(\bm{p} + \chi\bm{c}\Big) , \nn && \bm{\nabla}_{\bm{p}} \times \boldsymbol{\mathcal{E}} - \frac{1}{\mathcal{C}}\partial_{t} \boldsymbol{\mathcal{B}} = -\frac{2\pi}{\mathcal{C}} \sum_{\chi} \dot{\bm c} \delta\Big(\bm{p} + \chi\bm{c}\Big) , \nn && \bm{\nabla}_{\bm{p}} \times \boldsymbol{\mathcal{B}} + \frac{1}{\mathcal{C}}\partial_{t} \boldsymbol{\mathcal{E}} = 0 , ~~~~~ \bm{\nabla}_{\bm{p}} \cdot \boldsymbol{\mathcal{E}} = 0 . \label{Berry_Maxwell_Equation} \eqa
Here, both Berry magnetic and electric fields are given by the sum of all chiral charges
\bqa && \boldsymbol{\mathcal{B}} = \sum_{\chi} \boldsymbol{\mathcal{B}}_{\chi} , ~~~~~ \boldsymbol{\mathcal{E}} = \sum_{\chi} \boldsymbol{\mathcal{E}}_{\chi} . \eqa
The vector field $\bm{c}$ in momentum space corresponds to the distance between a pair of Weyl points, given by Eq. (\ref{Chiral_Gauge_Field}). The chirality is identified with a magnetic monopole in momentum space. The right hand side in the second equation of the Berry-Maxwell equation Eq. (\ref{Berry_Maxwell_Equation}) describes a monopole current in momentum space. The third and fourth equations may be regarded to be vector-field identities, referred to as the Bianchi identity \cite{Ryder_QFT_Textbook}. Here, we do not find quantities that correspond to the electrical current and charge density of Maxwell equations. An interesting quantity $\mathcal{C}$ is proposed to play the role of the speed of light in momentum space, i.e., the propagating speed of Berry electromagnetic waves in momentum space. Derivation of the speed of the Berry electromagnetic wave is on progress.

In order to figure out Berry-Maxwell equations, we start from a Lorentz invariant solution for the Berry magnetic field, given by \cite{Jackson_EM_Textbook}
\bqa \bm{\mathfrak{B}}_{\chi} = \frac{\chi}{2}\frac{\bm{p}+\chi\bm{c}}{|\bm{p}+\chi\bm{c}|^3 \gamma^2 [1-\beta^2\sin^2\Psi_{\chi}]^{3/2}} . \label{eq: Bchi} \eqa
where
\begin{gather}
 \cos\Psi_{\chi} = - \chi \frac{\bm{\dot{c}}}{|\bm{\dot{c}}|} \cdot\frac{\bm{p}+\chi\bm{c}}{|\bm{p}+\chi\bm{c}|} \\
 \gamma = \frac{1}{\sqrt{1-\beta^2}}, \;\; \beta=\frac{|\dot{\bm{c}}|}{\mathcal{C}} .
\end{gather}
This expression is reduced into
\bqa \bm{\mathfrak{B}}_{\chi} \approx \bm{\mathcal{B}}_{\chi} + \delta \bm{\mathcal{B}}_{\chi} , \eqa
taking into account $\beta \ll 1$. We obtain $\bm{\mathcal{B}}_{\chi} = \frac{\chi}{2} \frac{\bm{p}+\chi\bm{c}}{|\bm{p}+\chi\bm{c}|^3}$ in the $\mathcal{O}(\beta^{0})$ order and $\delta \bm{\mathcal{B}}_{\chi} = \beta^2 \Big( \frac{3}{2} \sin^2 \Psi_{\chi} - 1 \Big) \bm{\mathcal{B}}_{\chi}$ in the $\mathcal{O}(\beta^{2})$ order. There do not appear any Berry magnetic fields in the $\mathcal{O}(\beta^{1})$ order.

We introduce the $\mathcal{O}(\beta^{1})$ order as follows
\bqa && \boldsymbol{\mathcal{E}}_{\chi} \approx -\frac{1}{\mathcal{C}} \chi\dot{\bm c} \times \boldsymbol{\mathcal{B}}_{\chi } , \label{Berry_Electric_Field} \eqa
identified with Berry electric field. Then, the curl of the Berry electric field Eq. (\ref{Berry_Electric_Field}) in the second equation of Eq. (\ref{Berry_Maxwell_Equation}) is given by
\bqa && \bm{\nabla}_{\bm{p}} \times \boldsymbol{\mathcal{E}}_{\chi} =- \chi \frac{\dot{\bm c}}{\mathcal{C}} \bm{\nabla}_{\bm{p}} \cdot \boldsymbol{\mathcal{B}}_{\chi  } +\chi \frac{\dot{\bm c} }{\mathcal{C}}\cdot \bm{\nabla}_{\bm{p}} \boldsymbol{\mathcal{B}}_{\chi} \eqa
in the $\mathcal{O}(\beta^{1})$ order. The time derivative of the Berry magnetic field Eq. (\ref{eq: Bchi}) is
\bqa && \frac{1}{\mathcal{C}} \partial_{t} \bm{\mathfrak{B}}_{\chi} = \chi\frac{ \dot{\bm c}}{\mathcal{C}} \cdot \bm{\nabla}_{\bm{p}} \boldsymbol{\mathcal{B}}_{\chi } + \mathcal{O}(\beta^3) . \eqa
As a result, we confirm that the second equation of Eq. (\ref{Berry_Maxwell_Equation}) holds up to the $\mathcal{O}(\beta^{1})$ order.

In order to check out the third equation of Eq. (\ref{Berry_Maxwell_Equation}), we consider $\ddot{\bm{c}}=0$ for simplicity and $\dot{\bm{c}}=|\dot{\bm{c}}|\hat{z}$ without loss of generality. The curl of the Berry magnetic field Eq. (\ref{eq: Bchi}) is given by
\bqa && \nabla_{\bm{p}} \times \bm{\mathfrak{B}}_{\chi} = \nabla_{\bm{p}} \times \delta \bm{\mathcal{B}}_{\chi} \nn && = - \frac{3}{2}\beta^2\chi\frac{p_z+\chi c_z}{|\bm{p}+\chi\bm{c}|^5}
\Big[(p_x+\chi c_x)\hat{y}-(p_y+\chi c_y)\hat{x}\Big] . \nn \eqa
The time derivative of the Berry electric field Eq. (\ref{Berry_Electric_Field}) is
\bqa && \frac{1}{\mathcal{C}}\partial_t \mathcal{E}_{\chi}=\frac{3}{2}\beta^2\chi\frac{p_z+\chi c_z}{|\bm{p}+\chi\bm{c}|^5} \Big[(p_x+\chi c_x)\hat{y}-(p_y+\chi c_y)\hat{x}\Big] . \nn \eqa
As a result, we find that the third equation of the Berry-Maxwell equation Eq. (\ref{Berry_Maxwell_Equation}) holds up to the $\mathcal{O}(\beta^{2})$ order.

\section{Derivation of the topologically modified Drude model}

In order to prove the topologically modified Drude model, in particular, the emergence of the Berry electric field, we start from an effective Hamiltonian for a Weyl metal phase, given by
\bqa && H_\chi = \chi {\bm \sigma} \cdot \Big({\bm p} + \chi \bm{c} + \frac{e}{c} \bm{A} \Big) + e \Phi . \eqa
$\chi = \pm$ is a chiral charge. ${\bm \sigma}$ is a two-by-two Pauli matrix. $\bm{p}$ is a momentum. $\bm{c}$ is a chiral gauge field, given by Eq. (\ref{Chiral_Gauge_Field}). $\bm{A}$ and $\Phi$ are electromagnetic vector and scalar potentials with an electric charge $e$ and the speed of light $c$. This effective Hamiltonian gives rise to the transition amplitude
\bqa
&& \langle f | e^{-iH_\chi (t_f - t_i)} | i \rangle = \int_{\bm{r}_{i}}^{\bm{r}_{f}} D \bm{r} \int D \bm{p} \ \exp \bigg[ i \int_{t_i}^{t_f} dt \Big\{ {\bm p} \cdot \dot{\bm r} \nn && - \chi {\bm \sigma} \cdot \Big({\bm p} +\chi\bm{c} + \frac{e}{c} \bm{A} \Big) - e \Phi \Big\} \bigg] ,
\eqa
where $\hbar = 1$.

In order to describe low energy dynamics of electrons near a Fermi surface, we do not need to know the information of high energy electrons deep inside the Fermi surface, generally speaking. However, we are not allowed to neglect high energy dynamics of electrons in a Weyl metal phase when we deal with a pair of chiral Fermi surfaces. In particular, the topological information involved with the pair of Weyl points should be taken into account, integrating over such high energy electrons. In order to understand the low energy dynamics near a pair of chiral Fermi surfaces, we should integrate over high energy electron fields near the pair of Weyl points.

The integration of high energy electrons can be performed, rewriting the effective Weyl Hamiltonian in terms of a diagonalized basis,
\bqa && U_{\tilde{\bm p}}^{\dagger} {\bm \sigma} \cdot \tilde{\bm p} U_{\tilde{\bm p}} = | \tilde{\bm p} | \sigma^3 , \eqa
where $\tilde{\bm p} \equiv \bm p +\chi\bm c$. $U_{\tilde{\bm p}}$ is a two-by-two unitary matrix, expressed by $U_{\tilde{\bm p}} = \left( u_{\tilde{\bm p}} \ v_{\tilde{\bm p}} \right)$, where two-component column vectors are determined by
\bqa && ({\bm \sigma} \cdot \tilde{\bm p}) u_{\tilde{\bm p}} = |\tilde{\bm p}| u_{\tilde{\bm p}}, \ \ \ ({\bm \sigma} \cdot \tilde{\bm p}) v_{\tilde{\bm p}} = -|\tilde{\bm p}| v_{\tilde{\bm p}} , \eqa
respectively.

In order to describe the low energy dynamics of electrons near a pair of chiral Fermi surfaces, we neglect off-diagonal terms and take the $11-$component for $\chi = +$ and $22-$component for $\chi=-$. In other words, we consider \cite{CME4,CME6}
\bqa && \left( U_{\tilde{\bm p}'}^\dagger \exp \bigg[ -i{\bm \sigma} \cdot \Big( \tilde{\bm p} + \frac{e}{c} {\bm A} \Big) \Delta t \bigg] U_{\tilde{\bm p}} \right)_{11} \nn
&&\approx u_{\tilde{\bm p}'}^\dagger u_{\tilde{\bm p}} \exp \bigg[ -i \frac{|\tilde{\bm p}| + |\tilde{\bm p}'|}{2} \Delta t - i \frac{e}{c} \frac{\hat{\tilde{\bm p}} + \hat{\tilde{\bm p}}'}{2} \cdot \bm A \Delta t \nn &&\hspace{100pt} - \frac{e}{c} \frac{\Delta\tilde{\bm p} \times \hat{\tilde{\bm p}}}{2|\tilde{\bm p}|} \cdot \bm A \Delta t \bigg] \nn
&&\approx u_{\tilde{\bm p}'}^\dagger u_{\tilde{\bm p}} \exp \bigg[ -\frac{i}{2} \left( \Big| \tilde{\bm p} + \frac{e}{c}\bm A \Big| + \Big| \tilde{\bm p}' + \frac{e}{c}\bm A \Big| \right) \Delta t \nn &&\hspace{130pt} - i\frac{e}{c} \frac{\bm B \cdot \hat{\tilde{\bm p}}}{2|\tilde{\bm p}|} \Delta t \bigg] \nn
&&\approx u_{\tilde{\bm p}'}^\dagger u_{\tilde{\bm p}} \exp \bigg[ -i \left( 1 + \frac{e}{c} \bm B \cdot \bm{\mathcal{B}_{\bm p}}^+ \right) |\tilde{\bm p}| \Delta t \bigg] \eqa
in the path-integral representation, where $\hat{\tilde{\bm p}} \equiv \tilde{\bm p} /|\tilde{\bm p}|$ and $\bm{\mathcal{B}_{\bm p}}^+ = \frac{\hat{\tilde{\bm p}}}{2|\tilde{\bm p}|}$. Here, we used the Gordon identity, given by
\bqa u_{\bm p_\chi'}^\dagger \sigma^i u_{\bm p_\chi} = u_{\bm p_\chi'}^\dagger \bigg[ \frac{p_\chi^i + {p_\chi'}^i}{|\bm p_\chi| + |\bm p_\chi'|} - \frac{i\epsilon_{ijk} \Delta p_\chi^j }{|\bm p_\chi| + |\bm p_\chi'|} \sigma^k \bigg] u_{\bm p_\chi}, \eqa
up to the linear order in $\Delta p_\chi$. We also assume the semiclassical regime that the magnetic field is small enough for us to neglect the Landau level splitting. The Berry gauge field appears from
\bqa u_{\tilde{\bm p} + \Delta \tilde{\bm p}}^{\dagger} u_{\tilde{\bm p}} &&\approx \exp \big(-i \bm{\mathcal{A}}_{\bm p}^+ \cdot \Delta \tilde{\bm p} \big) \nn &&= \exp \bigg[ -i \bm{\mathcal{A}}_{\bm p}^+ \cdot \Big( \Delta \bm p +  \Delta \bm c \Big) \bigg] , \eqa
represented by
\bqa && \bm{\mathcal A}_{\bm p}^+ = i u_{\tilde{\bm p}}^\dagger \bm{\nabla}_{\bm p} u_{\tilde{\bm p}} . \eqa

Similarly, we find
\bqa && \left( U_{\tilde{\bm p}'}^\dagger \exp \bigg[ i{\bm \sigma} \cdot \Big( \tilde{\bm p} + \frac{e}{c} {\bm A} \Big) \Delta t \bigg] U_{\tilde{\bm p}} \right)_{22} \nn
&&\approx v_{\tilde{\bm p}'}^\dagger v_{\tilde{\bm p}} \exp \bigg[ -i \left( 1 + \frac{e}{c} \bm B \cdot \bm{\mathcal{B}_{\bm p}}^- \right) |\tilde{\bm p}| \Delta t \bigg] \eqa
with $\bm{\mathcal{B}_{\bm p}}^- = -\frac{\hat{\tilde{\bm p}}}{2|\tilde{\bm p}|}$. The Berry gauge field at $\chi = -$ results from
\bqa v_{\tilde{\bm p} + \Delta \tilde{\bm p}}^{\dagger} v_{\tilde{\bm p}} && = \exp \bigg[ -i \bm{\mathcal{A}}_{\bm p}^- \cdot \Big( \Delta \bm p -  \Delta \bm c \Big) \bigg] , \eqa
given by
\bqa && \bm{\mathcal A}_{\bm p}^- = i v_{\tilde{\bm p}}^\dagger \bm{\nabla}_{\bm p} v_{\tilde{\bm p}} =i u_{-\tilde{\bm p}}^\dagger \bm \nabla_{\bm p} u_{-\tilde{\bm p}} . \eqa

An effective action for the low energy dynamics of chiral fermions in a Weyl metal phase reads
\bqa && \mathcal{S}_{\chi}^{eff} = \int_{t_i}^{t_f} dt \Big\{ \Big({\bm p} + \frac{e}{c} \bm{A} \Big) \cdot \dot{\bm r} - e \Phi - \bm{\mathcal{A}}_{\bm p}^\chi \cdot \Big( \dot{\bm p} + \chi \dot{\bm c} \Big) \nn && - \Big( 1 + \frac{e}{c} {\bm B} \cdot \bm{\mathcal{B}}_{\bm p}^\chi \Big) \Big|{\bm p} + \chi\bm{c}\Big| \Big\} , \label{Effective_Weyl_Action_Chiral_Fermi_Surface}
\eqa
where
\bqa && \bm{\mathcal A}_{\bm p}^\chi = i u_{\chi\tilde{\bm p}}^\dagger \bm{\nabla}_{\bm p} u_{\chi\tilde{\bm p}} \eqa
is the Berry gauge field, originating from the high energy dynamics near the Weyl point. It is trivial to check out $\bm{\mathcal{B}}_{\bm p}^\chi = {\bm \nabla}_{\bm p} \times \bm{\mathcal{A}}_{\bm p}^\chi$. Here, a novel ingredient beyond the previous study is a coupling term $- \chi  \bm{\mathcal{A}}_{\bm p}^\chi \cdot \dot{\bm c}$.

It is straightforward to read the low energy effective Hamiltonian for a pair of chiral Fermi surfaces from the effective action Eq. (\ref{Effective_Weyl_Action_Chiral_Fermi_Surface}) as follows
\bqa && \mathcal{H}_{\chi} = - \frac{e}{c} \bm{A} \cdot \dot{\bm r} + e \Phi + \bm{\mathcal{A}}_{\bm p}^\chi \cdot \Big( \dot{\bm p} +\chi \dot{\bm c} \Big) \nn && + \Big( 1 + \frac{e}{c} {\bm B} \cdot \bm{\mathcal{B}}_{\bm p}^\chi \Big) \Big|{\bm p} +\chi \bm{c}\Big| . \eqa
Hamiltonian equations of motion $\dot{\bm r}_{\chi} = \frac{\partial \mathcal{H}_{\chi}}{\partial \bm{p}_\chi}$ and $\dot{\bm p}_{\chi} = - \frac{\partial \mathcal{H}_{\chi}}{\partial \bm{r}_\chi}$ give rise to the topologically modified Drude model Eqs. (\ref{Effective_Velocity_Weyl_Metal}) and (\ref{Effective_Force_Weyl_Metal}) with an emergent Berry electric field, where
\bqa \bm r \rightarrow \bm r_\chi, \ \ \ \bm p \rightarrow \bm p_\chi \eqa have been considered. The group velocity in Eq. (\ref{Effective_Velocity_Weyl_Metal}) is given by
\bqa && {\bm v}_{\chi} = {\bm \nabla}_{\bm p} \Big\{ \Big( 1 + \frac{e}{c} {\bm B} \cdot \bm{\mathcal{B}}_{\bm p}^\chi \Big) \Big|{\bm p} + \chi \bm{c}\Big| \Big\} . \eqa
This completes the derivation of the topologically modified Drude model.

\section{Current conservation law and chiral anomaly}

Solutions of the topological Drude model are given by
\bqa G_{\chi} \bm{\dot{p}}_{\chi} &=& e \bm{E} + \frac{e}{c} \Big( {\bm v}_{\chi} +\chi \dot{\bm c} \times \boldsymbol{\mathcal{B}}_{\chi} \Big) \times \bm{B} + \frac{e^{2}}{c} (\bm{E} \cdot \bm{B}) \boldsymbol{\mathcal{B}}_{\chi} , \nn G_{\chi} \bm{\dot{r}}_{\chi} &=& {\bm v}_{\chi} + \chi \dot{\bm c} \times \boldsymbol{\mathcal{B}}_{\chi} + e \bm{E} \times \boldsymbol{\mathcal{B}}_{\chi} + \frac{e}{c} ( \bm{v}_{\chi} \cdot \boldsymbol{\mathcal{B}}_{\chi} ) \bm{B} . \eqa
It is interesting to observe the symmetric structure between these two solutions, where the correspondence are
\bqa && e \bm{E} \longleftrightarrow {\bm v}_{\chi} +\chi \dot{\bm c} \times \boldsymbol{\mathcal{B}}_{\chi} , \nn && \frac{e}{c} \bm{B} \longleftrightarrow \boldsymbol{\mathcal{B}}_{\chi} . \eqa

Inserting these solutions into the Boltzmann equation Eq. (\ref{Boltzmann_Equation}), it is straightforward to find the current conservation equation
\bqa \partial_{t} \rho_{\chi} + \bm{\nabla}_{\bm{r}} \cdot \bm{j}_{\chi} &=& e\int \frac{d^{3} \bm{p}}{(2\pi)^{3}} (\partial_{t} G_{\chi}) f_{\chi} +\frac{\chi}{4\pi^2} \frac{e^{3}}{c} (\bm{E} \cdot \bm{B}) \nn &+& \frac{e^2}{c}\chi \int \frac{d^{3} \bm{p}}{(2\pi)^{3}}  [\bm{\nabla}_{\bm{p}} \cdot \{(\dot{\bm c} \times \boldsymbol{\mathcal{B}}_{\chi}) \times \bm{B}\} f_{\chi}] , \nn \eqa
where both electric charge density and current density are defined in Eq. (\ref{Density_Current_Weyl_Metal}).

The first term in the right hand side is
\bqa
&& e\int\frac{d^3p}{(2\pi)^3}(\partial_t G_{\chi})f_{\chi} = \frac{e^2}{c}\chi\int \frac{d^3p}{(2\pi)^3}f_{\chi}\dot{\bm{c}}\cdot\nabla_{\bm{p}}(\bm{B}\cdot\bm{\mathcal{B}}_{\chi}) . \nn
\eqa
The third term in the right hand side is
\bqa
&& \frac{e^2}{c} \chi \int \frac{d^{3} \bm{p}}{(2\pi)^{3}} [\bm{\nabla}_{\bm{p}} \cdot \{(\dot{\bm c} \times \boldsymbol{\mathcal{B}}_{\chi}) \times \bm{B}\} f_{\chi}] \nn && = \frac{1}{4\pi^2} \frac{e^2}{c} \bm{B}\cdot\dot{\bm{c}} - \frac{e^2}{c}\chi\int \frac{d^{3} \bm{p}}{(2\pi)^{3}} f_{\chi}\dot{\bm{c}}\cdot\nabla_{\bm{p}}(\bm{B}\cdot\bm{\mathcal{B}}_{\chi}) . \nn
\eqa
As a result, we obtain a modified current conservation law, given by
\begin{align}
\partial_t \rho_\chi+\nabla_{\bm{r}}\cdot\bm{j}_{\chi}=\frac{\chi}{4\pi^2}\frac{e^3}{c}\bm{E}\cdot\bm{B}+\frac{1}{4\pi^2}\frac{e^2}{c}\bm{B}\cdot\dot{\bm{c}} .
\end{align}

%
%\bqa && \rho_{\chi} = e\int \frac{d^{3} \bm{p}}{(2\pi)^{3}} G_{\chi} f_{\chi} , \nn && \bm{j}_{\chi} = e \int \frac{d^{3} \bm{p}}{(2\pi)^{3}} \Big\{ {\bm v}_{\chi} + \chi\dot{\bm c} \times \boldsymbol{\mathcal{B}}_{\chi} \nn && + e \bm{E} \times \boldsymbol{\mathcal{B}}_{\chi} + \frac{e}{c} ( \bm{v}_{\chi} \cdot \boldsymbol{\mathcal{B}}_{\chi} ) \bm{B} \Big\} f_{\chi} \eqa
%

Actually, this current conservation law has been known for a long time in the context of quantum field theory. An effective action for a Weyl metal phase is given by
\bqa && \mathcal{S}_{WM} = \int d^{4} x \bar{\psi} i \gamma_{\mu} [ \partial_{\mu} - i A_{\mu} - i \gamma_{5} c_{\mu}(t) ] \psi . \eqa
Here, $\psi$ is a four-component Dirac spinor. $\gamma^{\mu}$ is a four-by-four Dirac gamma matrix to satisfy the Clifford algebra with $\mu=0,1,2,3$ and $x^{\mu}=(t,x,y,z)$. $\gamma_{5}$ is a chiral matrix.

%
%$g^{\mu\nu}=-\delta^{\mu\nu}$ $\gamma^{4}=-i\gamma^0$, $A_4=iA_0$, $c_4=ic_0$
%
%\bqa && \partial_{\mu} \langle \bar{\psi} \gamma^{\mu} \gamma_{5} \psi\rangle =-i \frac{\epsilon^{\mu\nu\alpha\beta}}{16\pi^2}(F_{\mu\nu}F_{\alpha\beta}+f_{\mu\nu}f_{\alpha\beta})\eqa
%\bqa && \partial_{\mu}\langle\bar{\psi}\gamma^{\mu}\psi\rangle=-i\frac{\epsilon^{\mu\nu\alpha\beta}}{16\pi^2}(F_{\mu\nu}f_{\alpha\beta}+f_{\mu\nu}F_{\alpha\beta}) \eqa
%where $F_{\mu\nu}=\partial_{\mu} A_{\nu}-\partial_\nu A_{\mu}$, $f_{\mu\nu}=\partial_{\mu}c_\nu-\partial_\nu c_\mu$. Under Wick rotation($\tau\rightarrow-it$),
%\begin{gather}\partial_\tau\rightarrow i\partial_t\\A_4\rightarrow iA_0\\c_4\rightarrow ic_0\end{gather}
%

Based on the Fuzikawa's path-integral method \cite{Peskin_Schroeder}, one can find modified anomaly equations in the presence of the chiral gauge field
\bqa
\partial_{\mu} \langle \bar{\psi} \gamma^{\mu} \gamma_{5} \psi\rangle &=& \frac{1}{16\pi^2} \epsilon^{\mu\nu\alpha\beta} (F_{\mu\nu}F_{\alpha\beta}+f_{\mu\nu}f_{\alpha\beta}) \nn &=& \frac{1}{2\pi^2}\bm{E}\cdot\bm{B}
\eqa
and
\bqa
\partial_{\mu}\langle\bar{\psi}\gamma^{\mu}\psi\rangle &=& -\frac{1}{16\pi^2} \epsilon^{\mu\nu\alpha\beta} (F_{\mu\nu}f_{\alpha\beta}+f_{\mu\nu}F_{\alpha\beta}) \nn &=& \frac{1}{2\pi^2}\dot{\bm{c}}\cdot\bm{B} , \label{Covariant_Current_Anomaly}
\eqa
where $F_{\mu\nu}=\partial_{\mu} A_{\nu}-\partial_\nu A_{\mu}$ and $f_{\mu\nu}=\partial_{\mu}c_\nu-\partial_\nu c_\mu$ are field strength tensors for electromagnetic and chiral gauge fields, respectively \cite{Veltman_QFT_Textbook,Qi_Chiral_Gauge_Field_Anomaly}. Here, $\langle \bar{\psi} \gamma^{\mu} \gamma_{5} \psi\rangle = \bm{j}_{+} - \bm{j}_{-}$ is the chiral current and $\langle \bar{\psi} \gamma^{\mu} \psi\rangle = \bm{j}_{+} + \bm{j}_{-}$ is the electromagnetic U(1) current, where $\bm{j}_{\chi}$ is given by Eq. (\ref{Density_Current_Weyl_Metal}). We recall that the chiral gauge field is $(0,\bm{c}(t))$, resulting in the last equation Eq. (\ref{Covariant_Current_Anomaly}). Since the electromagnetic U(1) current is not conserved, one may suspect the validity of this result. However, this originates from the definition of the electromagnetic current. This chiral anomaly is known to be the anomaly of a covariant current \cite{Covariant_vs_Conserved_Current_Anomaly}.

\section{Role of the emergent Berry electric field in transport}

An essential question is on the role of the Berry electric field in anomalous transport of a Weyl metal phase. Resorting to the framework of Boltzmann transport theory, we resolve this issue. Solving the Boltzmann equation, we obtain the distribution function up to the linear order in the electric field as follows
\begin{align}
&f_{\chi}(t,\bm{p}) \approx f_{\chi}^{eq}(\bm{p}) \nonumber\\
 & - \frac{\partial f_{\chi}^{eq}(\varepsilon)}{\partial \varepsilon}  \Big( \int_{-\infty}^{\infty} d \omega e^{i \omega t} \frac{\tau_{eff}}{1 - i \omega \tau_{eff}} \bm{\dot{p}}_{\chi}(\omega) \Big) \cdot \bm{v}_{\chi}(t)
 \end{align}
 where
 \begin{align}
f_{\chi}^{eq}(\mathbf{p})&=\frac{1}{e^{\beta(G_{\chi}(\bm{p})|\mathbf{p}+\chi\mathbf{c}|-\mu)}+1},\\
\frac{\partial f_{\chi}^{eq}(\bm{p})}{\partial \epsilon}&\equiv \frac{\partial}{\partial \epsilon}\Big(\frac{1}{e^{\beta(\epsilon-\mu)}+1}\Big)\Big|_{\epsilon=G_{\chi}(\bm{p})|\bm{p}+\chi\bm{c}|}.
\end{align}
Here, $f_{\chi}^{eq}(\bm{p})$ is an equilibrium distribution function, and its derivative with respect to energy is $\frac{\partial f_{\chi}^{eq}(\bm{p})}{\partial \epsilon}$.
Then, the electric current reads
\begin{align}
\bm{j}_{\chi} &= e \int \frac{d^{3} \bm{p}}{(2\pi)^{3}}G_{\chi} \bm{\dot{r}}_{\chi} f_{\chi}^{eq}(\bm{p}) + e \int \frac{d^{3} \bm{p}}{(2\pi)^{3}} \Big( - \frac{\partial f_{\chi}^{eq}(\varepsilon)}{\partial \varepsilon} \Big) \nonumber\\
& \times G_{\chi} \bm{\dot{r}}_{\chi} \Big( \int_{-\infty}^{\infty} d \omega e^{i \omega t} \frac{\tau_{eff}}{1 - i \omega \tau_{eff}} \bm{\dot{p}}_{\chi}(\omega) \Big) \cdot \bm{v}_{\chi}(t) .
\end{align}

%
%\bqa && \bm{j}_{\chi} = e \int \frac{d^{3} \bm{p}}{(2\pi)^{3}} \Big\{ {\bm v}_{\chi} + \frac{\chi}{\mathcal{C}} \dot{\bm c} \times \boldsymbol{\mathcal{B}}_{\chi} + e \bm{E} \times \boldsymbol{\mathcal{B}}_{\chi} \nn && + \frac{e}{c} ( \bm{v}_{\chi} \cdot \boldsymbol{\mathcal{B}}_{\chi} ) \bm{B} \Big\} f_{eq}(\bm{p}) + e \int \frac{d^{3} \bm{p}}{(2\pi)^{3}} \Big( - \frac{\partial f_{eq}(\varepsilon)}{\partial \varepsilon} \Big) \Big\{ {\bm v}_{\chi} \nn && + \frac{\chi}{\mathcal{C}} \dot{\bm c} \times \boldsymbol{\mathcal{B}}_{\chi} + e \bm{E} \times \boldsymbol{\mathcal{B}}_{\chi} + \frac{e}{c} ( \bm{v}_{\chi} \cdot \boldsymbol{\mathcal{B}}_{\chi} ) \bm{B} \Big\} \nn && \Big( \int_{-\infty}^{\infty} d \omega e^{i \omega t} \frac{\tau_{eff}}{1 - i \omega \tau_{eff}} \bm{\dot{p}}_{\chi}(\omega) \Big) \cdot \bm{v}_{\chi}(t) \eqa
%

In this study we focus on the adiabatic regime, defined by
\bqa && \omega \tau_{eff} \ll 1 . \eqa
Then, the distribution function can be simplified as
\bqa && f_{\chi}(t,\bm{p}) \approx f_{\chi }^{eq}(\bm{p}) + \tau_{eff} (\bm{\dot{p}}_{\chi}(t) \cdot \bm{v}_{\chi}(t)) \Big( - \frac{\partial f_{\chi }^{eq}(\varepsilon)}{\partial \varepsilon} \Big) . \nn \eqa
Accordingly, the electric current is given by
\bqa && \bm{j}_{\chi} = e \int \frac{d^{3} \bm{p}}{(2\pi)^{3}}G_{\chi} \bm{\dot{r}}_{\chi} f_{\chi }^{eq}(\bm{p}) \nn && + e \tau_{eff} \int \frac{d^{3} \bm{p}}{(2\pi)^{3}} \Big( - \frac{\partial f_{\chi eq}(\bm{p})}{\partial \varepsilon} \Big) G_{\chi} \bm{\dot{r}}_{\chi} \bm{\dot{p}}_{\chi}(t) \cdot \bm{v}_{\chi}(t) . \nn \eqa
As a result, we obtain an electrical current in a Weyl metal phase
\bqa && \bm{j} = \bm{j}_{AHE} + \bm{j}_{CME} + \bm{j}_{LMC} + \Delta \bm{j}_{BE} . \eqa
Here,
\bqa && \bm{j}_{AHE} = e^{2} \int \frac{d^{3} \bm{p}}{(2\pi)^{3}} \sum_{\chi = \pm} (\bm{E} \times \bm{\mathfrak{B}}_{\chi}) f_{\chi}^{eq}(\bm{p}) \label{Conventional_AHE} \eqa
describes an anomalous Hall effect, resulting from the Berry magnetic field \cite{AHE1,AHE2,AHE3,Berry_Curvature_Review1,Berry_Curvature_Review2}.
\bqa && \bm{j}_{CME} = \frac{e^{2}}{c} \int \frac{d^{3} \bm{p}}{(2\pi)^{3}} \sum_{\chi = \pm} ( \bm{v}_{\chi} \cdot \bm{\mathfrak{B}}_{\chi} ) \bm{B} f_{\chi}^{eq}(\bm{p}) \eqa
gives rise to the chiral magnetic effect \cite{CME1,CME2,CME3,CME4,CME5,CME6,CME7}, which can occur when the chiral chemical potential exists. These two types of currents are dissipationless in nature, where even high energy electrons deep inside a pair of chiral Fermi surfaces are involved.
\bqa && \bm{j}_{LMC} = e \tau_{eff} \int \frac{d^{3} \bm{p}}{(2\pi)^{3}} \Big( - \frac{\partial f_{\chi}^{eq}(\bm{p})}{\partial \varepsilon} \Big) G_{\chi}^{-1} \bm{v}_{\chi} \cdot \Big\{ e \bm{E} \nn && + \frac{e}{c} {\bm v}_{\chi} \times \bm{B} + \frac{e^{2}}{c} (\bm{E} \cdot \bm{B}) \bm{\mathfrak{B}}_{\chi} \Big\} \Big\{ {\bm v}_{\chi} + e \bm{E} \times \bm{\mathfrak{B}}_{\chi} \nn && + \frac{e}{c} ( \bm{v}_{\chi} \cdot \bm{\mathfrak{B}}_{\chi} ) \bm{B} \Big\} \eqa
results in the longitudinal negative magnetoresistivity, which occurs when the electric current is driven along the direction of the pair of Weyl points. Although this electric current is a Fermi-surface contribution, the chiral anomaly plays a central role in this transport coefficient, regarded to be a fingerprint of a Weyl metal phase \cite{NLMR_First_Exp,NLMR_Followup_I,NLMR_Followup_II,NLMR_Followup_III,Nielsen_Ninomiya_NLMR,Boltzmann_Chiral_Anomaly1,Boltzmann_Chiral_Anomaly2,Boltzmann_Chiral_Anomaly3,Boltzmann_Chiral_Anomaly4,Boltzmann_Chiral_Anomaly5,Boltzmann_Chiral_Anomaly6}.

Other current contributions turn out to result from the Berry electric field. They are classified in the following way
\begin{gather} \Delta \bm{j}_{BE} = \Delta \bm{j}_{BE}^{(1)} + \Delta \bm{j}_{BE}^{(2)} + \Delta \bm{j}_{BE}^{(3)} + \Delta \bm{j}_{BE}^{(4)} , \end{gather}
where
\begin{gather}
\Delta \bm{j}_{BE}^{(1)} = e \sum_{\chi}\int \frac{d^{3} \bm{p}}{(2\pi)^{3}} \chi \dot{\bm c} \times \bm{\mathfrak{B}}_{\chi} f_{\chi}^{eq}(\bm{p}) 
\end{gather}
is a dissipationless current, given by the whole electrons inside both chiral Fermi surfaces, and
\bqa && \Delta \bm{j}_{BE}^{(2)} = e \tau_{eff}\sum_{\chi} \int \frac{d^{3} \bm{p}}{(2\pi)^{3}} \Big( - \frac{\partial f_{\chi}^{eq}(\bm{p})}{\partial \varepsilon} \Big) G_{\chi}^{-1} \bm{v}_{\chi} \cdot \Big\{ e \bm{E} \nn && + \frac{e}{c} {\bm v}_{\chi} \times \bm{B} + \frac{e^{2}}{c} (\bm{E} \cdot \bm{B}) \bm{\mathfrak{B}}_{\chi} \Big\} \chi \dot{\bm c} \times \bm{\mathfrak{B}}_{\chi} , \eqa
\bqa && \Delta \bm{j}_{BE}^{(3)} = e \tau_{eff} \sum_{\chi} \int \frac{d^{3} \bm{p}}{(2\pi)^{3}} \Big( - \frac{\partial f_{\chi}^{eq}(\bm{p})}{\partial \varepsilon} \Big) \nn && G_{\chi}^{-1} \Big\{ {\bm v}_{\chi} + e \bm{E} \times \bm{\mathfrak{B}}_{\chi} + \frac{e}{c} ( \bm{v}_{\chi} \cdot \bm{\mathfrak{B}}_{\chi} ) \bm{B} \Big\} \nn && \frac{e}{c}\chi \Big( (\dot{\bm c} \times \bm{\mathfrak{B}}_{\chi}) \times \bm{B} \Big) \cdot \bm{v}_{\chi} , \eqa
and
\bqa && \Delta \bm{j}_{BE}^{(4)} = e \tau_{eff}\sum_{\chi} \int \frac{d^{3} \bm{p}}{(2\pi)^{3}} \Big( - \frac{\partial f_{\chi}^{eq}(\bm{p})}{\partial \varepsilon} \Big) \nn && G_{\chi}^{-1} \chi\dot{\bm c} \times \boldsymbol{\mathfrak{B}}_{\chi} \frac{e}{c} \Big( \chi (\dot{\bm c} \times \boldsymbol{\mathfrak{B}}_{\chi}) \times \bm{B} \Big) \cdot \bm{v}_{\chi} \eqa
are dissipative currents with the effective scattering time, given by chiral fermions near the pair of chiral Fermi surfaces. We emphasize that all these currents are proportional to $\dot{\bm c}$.

We observe symmetry properties under the variable change of $\bm{p}\rightarrow -\bm{p}$ as follows
\begin{gather}
f_{\chi}^{eq}(-\bm{p})=f_{-\chi}^{eq}(\bm{p}) \\
\frac{\partial f_{\chi}^{eq}(-\bm{p})}{\partial \epsilon}=\frac{\partial f_{-\chi}^{eq}(\bm{p})}{\partial \epsilon} \\
\boldsymbol{\mathfrak{B}}_{\chi}(-\bm{p})=\boldsymbol{\mathfrak{B}}_{-\chi}(\bm{p}) \\
G_{\chi}(-\bm{p})=G_{-\chi}(\bm{p}) \\
\bm{v}_{\chi}(-\bm{p})=-\bm{v}_{-\chi}(\bm{p}) .
\end{gather}
Applying these symmetry properties to electric currents driven by the Berry electric field, we find that many terms vanish identically.
First, we obtain $\Delta \bm{j}_{BE}^{(1)} = 0$ and $\Delta \bm{j}_{BE}^{(4)}=0$.
Both the second and third contributions are also simplified as follows
\begin{align}
\Delta \bm{j}_{BE}^{(2)} &= e \tau_{eff}\sum_{\chi} \int \frac{d^{3} \bm{p}}{(2\pi)^{3}} \Big( - \frac{\partial f_{\chi}^{eq}(\bm{p})}{\partial \varepsilon} \Big) G_{\chi}^{-1} \bm{v}_{\chi} \cdot \Big\{ e \bm{E} \nn
& + \frac{e^{2}}{c} (\bm{E} \cdot \bm{B}) \bm{\mathfrak{B}}_{\chi} \Big\} \chi \dot{\bm c} \times \bm{\mathfrak{B}}_{\chi}
\end{align}
and
\begin{align}
\Delta \bm{j}_{BE}^{(3)} &= e \tau_{eff} \sum_{\chi}\int \frac{d^{3} \bm{p}}{(2\pi)^{3}} \Big( - \frac{\partial f_{\chi}^{eq}(\bm{p})}{\partial \varepsilon} \Big) \nn & G_{\chi}^{-1}\bm{v}_{\chi}\cdot\Big\{(\bm{B}\cdot\dot{\bm{c}}) \bm{\mathfrak{B}}_{\chi} - (\bm{B} \cdot \bm{\mathfrak{B}}_{\chi})\dot{\bm{c}}\Big\} \nn & \frac{e^2}{c} \chi \bm{E} \times \bm{\mathfrak{B}}_{\chi} ,
\end{align}
respectively. In the following discussion we will consider these transport coefficients up to the first order in $\beta$, described by $\bm{\mathcal{B}}_{\chi}$ and $\bm{\mathcal{E}}_{\chi}$ introduced before.

%
%\begin{align}
%\nabla_{\bm{p}}G_{\chi}&=\frac{e}{c}\nabla_{\bm{p}}(\bm{B}\cdot\mathcal{B}_{\chi})\nonumber\\
%&=\frac{e}{c}\Big[(\bm{B}\cdot\nabla_{\bm{p}})\mathcal{B}_{\chi}+\bm{B}\times(\nabla_{\bm{p}}\times\mathcal{B}_{\chi})\Big]\nonumber\\
%&=\frac{e}{c}(\bm{B}\cdot\nabla_{\bm{p}})\mathcal{B}_{\chi}-\frac{e}{c}\bm{B}\times\frac{1}
%{\mathcal{C}}\partial_t \mathcal{E}_\chi\\
%&\approx \frac{e}{c}(\bm{B}\cdot\nabla_{\bm{p}})\mathcal{B}_{\chi} \;(\because \text{2nd term}\sim O(\beta^2))
%\end{align}
%
%\begin{align}
%\therefore G_{\chi}(\bm{p})^{-1}\bm{v}_{\chi}(\bm{p})&\approx G_{\chi}^{-1}\nabla_{\bm{p}}\Big[G_{\chi}(\bm{p})\Big|\bm{p}+\chi\bm{c}\Big|\Big]\nonumber\\
%&=G_{\chi}^{-1}\nabla_{\bm{p}}G_{\chi}\Big|\bm{p}+\chi\bm{c}\Big|+\frac{\bm{p}+\chi\bm{c}}{\Big|\bm{p}+\chi\bm{c}\Big|}\nonumber\\
%&=G_{\chi}^{-1}\Big|\bm{p}+\chi\bm{c}\Big|\frac{e}{c}(\bm{B}\cdot\nabla_{\bm{p}})\mathcal{B}_{\chi}+\frac{\bm{p}+\chi\bm{c}}{\Big|\bm{p}+\chi\bm{c}\Big|}\nonumber\\
%&\equiv \tilde{\bm{v}}_{\chi}
%\end{align}
%

In order to figure out the direction of electrical currents in terms of applied electric fields, magnetic fields, and the direction of $\dot{\bm{c}}$, we separate each vector quantity such as $\bm{E}$, $\bm{\mathcal{B}}_{\chi}$, $\dot{\bm{c}}$, and etc. into two components as follows: Parallel and perpendicular to $\bm{B}$, respectively,
\begin{gather}
 \bm{B}=\bm{B}_{||} , ~~~~~ \bm{c}=\bm{c}_{||} ~ (\because \bm{c} ~ || ~ \bm{B}) , \nn
\bm{E}=\bm{E}_{||}+\bm{E}_{\perp} , ~~~~~ \dot{\bm{c}}=\dot{\bm{c}}_{||}+\dot{\bm{c}}_{\perp} , \nn
\bm{\mathcal{B}}_{\chi}=\frac{\chi}{2}\frac{\bm{p}_{||}+\chi\bm{c}_{||}}{R_{\chi}(\theta)^3}+\frac{\chi}{2}\frac{\bm{p}_{\perp}}{R_{\chi}(\theta)^3} \equiv \bm{\mathcal{B}}_{\chi ||}(p, \theta)+\bm{\mathcal{B}}_{\chi\perp}(p,\theta,\phi) . \nn
\end{gather}
where
\begin{gather}
R_{\chi}(\theta) \equiv \Big|\bm{p}+\chi\bm{c}\Big| = \sqrt{\Big(p\cos\theta+\chi\bm{c}_{||}\Big)^2+p^2\sin^2\theta}
\end{gather}
with $\bm{p} = ( p\sin\theta\cos\phi,p\sin\theta\sin\phi,p\cos\theta)$. $\theta$ and $\phi$ are inclination and azimuth, respectively, in the polar coordinate, where $\hat{\bm{B}}_{||}$ is identified with $\hat{\bm{z}}$. Accordingly, we have
\begin{gather}
 f_{\chi}^{eq}(\bm{p})=f_{\chi}^{eq}(p,\theta), \nn
 \frac{\partial f_{\chi}^{eq}(\bm{p})}{\partial \epsilon}=\frac{\partial f_{\chi}^{eq}(p,\theta)}{\partial \epsilon},\nn
 G_{\chi}(\mathbf{p})=G_{\chi}(p,\theta)=1+\frac{e}{c}B_{||}\mathcal{B}_{\chi ||}(p,\theta).
 \end{gather}
 We introduce a modified group velocity as follows
 \begin{align}
 \tilde{\bm{v}}_{\chi}(\bm{p}) &\equiv G^{-1}_{\chi}(\bm{p})\bm{v}_{\chi}(\bm{p}) \nonumber \\
 &= G_{\chi}^{-1}\Big|\bm{p}+\chi\bm{c}\Big|\frac{e}{c}(\bm{B}\cdot\nabla_{\bm{p}})\bm{\mathcal{B}}_{\chi}+\frac{\bm{p}+\chi\bm{c}}{\Big|\bm{p}+\chi\bm{c}\Big|}
\end{align}
We also decompose this modified group velocity into
\bqa
\tilde{\bm{v}}_{\chi}&\equiv \tilde{\bm{v}}_{\chi ||}(p,\theta)+\tilde{\bm{v}}_{\chi\perp}(p,\theta,\phi) ,
\eqa
where
\bqa
\tilde{\bm{v}}_{\chi ||}(p,\theta) &=& \Big(G_{\chi}^{-1}(p,\theta)\chi|\bm{B}_{||}|\frac{e}{c}\frac{|\bm{p}_\perp|^2-2|\bm{p}_{||}+\chi\bm{c}_{||}|^2}{2R^4_{\chi}(p,\theta)|\bm{p}_{||}+\chi\bm{c}_{||}|}\nn &&+ \frac{1}{R_{\chi}(p,\theta)}\Big) \Big(\bm{p}_{||}+\chi\bm{c}_{||}\Big) , \nn
\tilde{\bm{v}}_{\chi \perp}(p,\theta) &=& \Big(-\frac{3}{2}G_{\chi}^{-1}(p,\theta)\chi |\bm{B}_{||}|\frac{e}{c}\frac{|\bm{p}_{||}+\chi\bm{c}_{||}|}{R^4_{\chi}(p,\theta)}\nn
&&+\frac{1}{R_{\chi}(p,\theta)}\Big) \bm{p}_\perp .
\eqa
As a result, both contributions for electrical currents are given by
\bqa
\Delta \bm{j}_{BE}^{(2)} &=& e\tau_{eff}\sum_{\chi} \int\frac{d^3p}{(2\pi)^3}\Big(-\frac{\partial f_{\chi}^{eq}(\bm{p})}{\partial \epsilon}\Big)\frac{e^2}{c}\chi \nn
&& \Big\{(\mathcal{B}_{\chi ||} {E}_{||})(\tilde{v}_{\chi ||} \mathcal{B}_{\chi ||}+\tilde{{v}}_{\chi\perp} \mathcal{B}_{\chi\perp}) (\dot{\bm{c}}_{\perp}\times\bm{B}_{||}) \nn
&+& \frac{c}{e}(\tilde{{v}}_{\chi ||} {E}_{||}) (\dot{\bm{c}}_{\perp}\times \bm{\mathcal{B}}_{\chi ||}) + \frac{c}{e}(\tilde{{v}}_{\chi\perp} {E}_\perp) (\dot{\bm{c}}_{||}\times\bm{\mathcal{B}}_{\chi\perp}) \nn
&+& \frac{e}{c}(\tilde{{v}}_{\chi \perp} {E}_{\perp}) (\dot{\bm{c}}_{\perp} \times \bm{\mathcal{B}}_{\chi\perp}) \Big\}
\eqa
and
\bqa
\Delta \bm{j}_{BE}^{(3)} &=& e\tau_{eff}\sum_{\chi}\int\frac{d^3p}{(2\pi)^3}\Big(-\frac{\partial f_{\chi}^{eq}(\bm{p})}{\partial \epsilon}\Big)\frac{e^2}{c}\chi \nn
&& \Big\{(\tilde{{v}}_{\chi\perp} \mathcal{B}_{\chi\perp})(\mathcal{B}_{\chi ||}\dot{{c}}_{||}) (\bm{E}_{\perp}\times \bm{B}_{||}) \nn
&-&(\tilde{{v}}_{\chi\perp}\dot{{c}}_\perp )({B}_{||}\mathcal{B}_{\chi ||}) (\bm{E}_{||} \times \bm{\mathcal{B}}_{\chi \perp}) \nn
&-&(\tilde{{v}}_{\chi\perp}\dot{{c}}_\perp )({B}_{||}\mathcal{B}_{\chi ||}) (\bm{E}_{\perp} \times \bm{\mathcal{B}}_{\chi \perp}) \Big\} ,
\eqa
where we have utilized the following properties
\begin{gather}
\bm{\mathcal{B}}_{\chi \perp}(p,\theta,\phi+\pi) = - \bm{\mathcal{B}}_{\chi\perp}(p,\theta,\phi) , \nn
\bm{\mathcal{B}}_{\chi ||}(p,\pi-\theta) = \bm{\mathcal{B}}_{-\chi ||}(p,\theta) , \nn
\bm{\mathcal{B}}_{\chi \perp}(p,\pi-\theta,\phi) = - \bm{\mathcal{B}}_{-\chi \perp}(p,\theta,\phi) , \nn
\tilde{\bm{v}}_{\chi\perp}(p,\theta,\phi+\pi) = - \tilde{\bm{v}}_{\chi\perp}(p,\theta,\phi) , \nn
\tilde{\bm{v}}_{\chi ||}(p,\pi-\theta,\phi) = -\tilde{\bm{v}}_{-\chi ||}(p,\theta,\phi) , \nn
\tilde{\bm{v}}_{\chi \perp}(p,\pi-\theta,\phi) = \tilde{\bm{v}}_{-\chi \perp}(p,\theta,\phi) , \nn
f_{\chi}^{eq}(p,\pi-\theta) = f_{-\chi}^{eq}(p,\theta) ,\;\;
\frac{\partial f_{\chi}^{eq}(p,\pi-\theta)}{\partial \epsilon} = \frac{\partial f_{-\chi}^{eq}(p,\theta)}{\partial \epsilon} , \nn
R_{\chi}(p,\pi-\theta) = R_{-\chi}(p,\theta) , \;\; G_{\chi}(\pi-\theta)=G_{-\chi}(\theta) .
\end{gather}

It is more convenient to reexpress these currents in the following way
\begin{align}
\Delta\bm{j}_{BE}^{(2)}+\Delta\bm{j}_{BE}^{(3)} \equiv \bm{j}_{B,E}+\bm{j}_{E} ,
\end{align}
considering the $\bm{B}_{||}$ dependence. Here, $\bm{j}_{B,E}$ and $\bm{j}_E$ are
\begin{align}
\bm{j}_{B,E}&=e\tau_{eff}\sum_{\chi}\int\frac{d^3p}{(2\pi)^3}\Big(-\frac{\partial f_{\chi}^{eq}(\bm{p})}{\partial \epsilon}\Big)\frac{e^2}{c}\chi\nonumber\\&\Big\{(\mathcal{B}_{\chi ||} {E}_{||})(\tilde{{v}}_{\chi ||}\mathcal{B}_{\chi ||}+\tilde{{v}}_{\chi\perp}\mathcal{B}_{\chi\perp}) (\dot{\bm{c}}_{\perp}\times\bm{B}_{||}) \nonumber\\
&+(\tilde{{v}}_{\chi\perp}\mathcal{B}_{\chi\perp})(\mathcal{B}_{\chi ||}\dot{{c}}_{||}) (\bm{E}_{\perp}\times \bm{B}_{||}) \nonumber\\
&-(\tilde{{v}}_{\chi\perp}\dot{{c}}_\perp )({B}_{||}\mathcal{B}_{\chi ||}) (\bm{E}_{||} \times \bm{\mathcal{B}}_{\chi \perp}) \nonumber\\
&-(\tilde{{v}}_{\chi\perp}\dot{{c}}_\perp )({B}_{||}\mathcal{B}_{\chi ||}) (\bm{E}_{\perp} \times \bm{\mathcal{B}}_{\chi \perp}) \Big\} \label{AHE_BE_Berry_Electric_Field}
\end{align}
and
\begin{align}
\bm{j}_{E}&=e\tau_{eff}\sum_{\chi}\int\frac{d^3p}{(2\pi)^3}\Big(-\frac{\partial f_{\chi}^{eq}(\bm{p})}{\partial \epsilon}\Big)\chi e\Big\{(\tilde{{v}}_{\chi ||}{E}_{||}) (\dot{\bm{c}}_{\perp}\times\bm{\mathcal{B}}_{\chi ||}) \nonumber\\ &+(\tilde{{v}}_{\chi\perp}{E}_\perp) (\dot{\bm{c}}_{||}\times\bm{\mathcal{B}}_{\chi\perp}) + (\tilde{{v}}_{\chi \perp}{E}_{\perp}) (\dot{\bm{c}}_{\perp} \times \bm{\mathcal{B}}_{\chi\perp}) \Big\} , \label{AHE_E_Berry_Electric_Field}
\end{align}
respectively.

Based on these equations, we determine the direction of an anomalous current associated with the Berry electric field. For simplicity, we assume
\begin{align}
\bm{B}(t)=\bm{B}_{0} + \bm{\delta B}(t)
\end{align}
with $|\bm{B}_{0}|\gg |\bm{\delta B}(t)|$. Note that $\dot{\bm{c}}\propto \delta\dot{\bm{B}}(t)$ since $\bm{c}\propto \bm{B}$.

First, we consider the case of $\bm{B}_0\parallel \bm{E}\parallel \dot{\bm{c}}$. Then, we obtain
\begin{align}
\bm{j}_{B,E}=\bm{j}_{E}=0 .
\end{align}

Second, we consider the case of $\bm{B}_0\parallel \bm{E}\perp \dot{\bm{c}}$. Then, we find
\begin{align}
\bm{j}_{B,E}&=e\tau_{eff}\sum_{\chi}\int\frac{d^3p}{(2\pi)^3}\Big(-\frac{\partial f_{\chi}^{eq}(\bm{p})}{\partial \epsilon}\Big)\frac{e^2}{c}\chi\nonumber\\&\Big\{(\mathcal{B}_{\chi ||} {E}_{||})(\tilde{{v}}_{\chi ||} \mathcal{B}_{\chi ||}+\tilde{{v}}_{\chi\perp} \mathcal{B}_{\chi\perp}) (\dot{\bm{c}}_{\perp}\times\bm{B}_0) \nonumber\\
&-(\tilde{{v}}_{\chi\perp}\dot{{c}}_\perp )({B}_0\mathcal{B}_{\chi ||}) (\bm{E}_{||} \times \bm{\mathcal{B}}_{\chi \perp}) \Big\} ,
\end{align}
parallel with $(\hat{\dot{\bm{c}}}\times\hat{\bm{E}})$, and
\begin{align}
\bm{j}_{E}&=e\tau_{eff}\sum_{\chi}\int\frac{d^3p}{(2\pi)^3}\Big(-\frac{\partial f_{\chi eq}(\bm{p})}{\partial \epsilon}\Big)\chi e\nonumber\\
& (\tilde{ {v}}_{\chi ||} {E}_{||}) (\dot{\bm{c}}_{\perp} \times \bm{\mathcal{B}}_{\chi ||}) ,
\end{align}
parallel with $(\hat{\dot{\bm{c}}}\times\hat{\bm{E}})$. We emphasize that these anomalous Hall currents are in the order of $\mathcal{O}[\{\bm{\delta B}(t)\}^{0}]$, which should be distinguished from ``conventional" anomalous Hall currents in the order of $\mathcal{O}[\{\bm{\delta B}(t)\}^{1}]$. These are novel anomalous Hall currents beyond the Berry magnetic field.

Third, we consider the case of $\bm{B}_0 \perp \bm{E} || \dot{\bm{c}}$. Then, we have
\begin{align}
\bm{j}_{B,E}=\bm{j}_{E}=0 .
\end{align}
Of course, there exists a ``conventional" anomalous Hall current, described by Eq. (\ref{Conventional_AHE}), since the applied electric field $\bm{E}$ is orthogonal to the applied magnetic field $\bm{B}_0$.

Fourth, we consider the case of $\bm{B}_0 \perp \bm{E}$ with $\bm{B}_0 || \dot{\bm{c}}$. Then, we obtain
\begin{align}
\bm{j}_{B,E}&=e\tau_{eff}\sum_{\chi}\int\frac{d^3p}{(2\pi)^3}\Big(-\frac{\partial f_{\chi}^{eq}(\bm{p})}{\partial \epsilon}\Big)\frac{e^2}{c}\chi\nonumber\\
& (\tilde{{v}}_{\chi\perp}\mathcal{B}_{\chi\perp})(\mathcal{B}_{\chi ||}\dot{{c}}_{||}) (\bm{E}_{\perp}\times \bm{B}_0) ,
\end{align}
parallel with $(\hat{E}\times\hat{\dot{\bm{c}}})$, and
\begin{align}
\bm{j}_{E}&=e\tau_{eff}\sum_{\chi}\int\frac{d^3p}{(2\pi)^3}\Big(-\frac{\partial f_{\chi}^{eq}(\bm{p})}{\partial \epsilon}\Big)\chi e\nonumber\\
& (\tilde{{v}}_{\chi\perp}{E}_\perp) (\dot{\bm{c}}_{||}\times\bm{\mathcal{B}}_{\chi\perp}) ,
\end{align}
parallel with $-(\hat{E}\times\hat{\dot{\bm{c}}})$. Here, we also point out the existence of the ``conventional" anomalous Hall current, which does not depend on $\dot{\bm{c}}_{||}$. These anomalous Hall currents are driven by the emergent Berry electric field.

Finally, we consider the case of $\bm{B}_0 \perp \bm{E}$, $\bm{E}\perp \dot{\bm{c}}$, and $\bm{B}_0\perp\dot{\bm{c}}$. Then, we obtain
\begin{align}
\bm{j}_{B,E}&=e\tau_{eff}\sum_{\chi}\int\frac{d^3p}{(2\pi)^3}\Big(-\frac{\partial f_{\chi}^{eq}(\bm{p})}{\partial \epsilon}\Big)\frac{e^2}{c}\chi\nonumber\\
& \{- (\tilde{{v}}_{\chi\perp}\dot{{c}}_\perp )({B}_0\mathcal{B}_{\chi ||}) (\bm{E}_{\perp} \times \bm{\mathcal{B}}_{\chi \perp})\} ,
\end{align}
parallel with $(-\hat{\bm{B}}_0)$, and
\begin{align}
\bm{j}_{E}&=e\tau_{eff}\sum_{\chi}\int\frac{d^3p}{(2\pi)^3}\Big(-\frac{\partial f_{\chi}^{eq}(\bm{p})}{\partial \epsilon}\Big)\chi e\nonumber\\
& (\tilde{{v}}_{\chi \perp}{E}_{\perp}) (\dot{\bm{c}}_{\perp} \times \bm{\mathcal{B}}_{\chi\perp}) \Big\} ,
\end{align}
parallel with $(-\hat{\bm{B}}_0)$. We recall that the conventional anomalous Hall current is given along the direction of $\bm{B}_{0} \times \bm{E}$. On the other hand, these anomalous Hall currents are driven in parallel with the applied magnetic field $\bm{B}_{0}$, quite surprising. 

All these situations are summarized in Table I and pictorially shown in Fig. 1.

\begingroup
\renewcommand{\arraystretch}{1.4}
\begin{table}[H]
\begin{ruledtabular}
\begin{tabular}{c|c|c|c}
&  & Direction of $\mathbf{j}_{E}$&Direction of $\mathbf{j}_{B,E}$\\
\hline
\multirow{2}{*}{$\bm{B}_0||\bm{E}$}&$\bm{B}_0||\dot{\mathbf{c}}$& $\cdot$ &$\cdot$  \\
\cline{2-4}
   & $\bm{B}_0\perp \dot{\mathbf{c}}$ & $\dot{\bm{c}}\times\bm{E}$&$\dot{\bm{c}}\times\bm{E}$\\
\hline
\multirow{3}{*}{$\bm{B}_0\perp \bm{E}$}&$\dot{\bm{c}}||\bm{E}$&$\cdot$ &$\cdot$ \\
\cline{2-4}
 &$\dot{\bm{c}}||\bm{B}_0$&$\dot{\bm{c}}\times\bm{E}$&$\dot{\bm{c}}\times\bm{E}$\\
 \cline{2-4}
 & $\dot{\bm{c}}\perp\bm{E} \; \& \; \dot{\bm{c}}\perp\bm{B}_0$ &$ \bm{B}_0$ & $\bm{B}_0$
\end{tabular}
\end{ruledtabular}
\caption{Anomalous Hall currents $\bm{j}_E$ [Eq. (\ref{AHE_E_Berry_Electric_Field})] and $\bm{j}_{B,E}$ [Eq. (\ref{AHE_BE_Berry_Electric_Field})] driven by the Berry electric field}
\end{table}
\endgroup

\begin{figure}[h]
\centering
\includegraphics[scale=0.35]{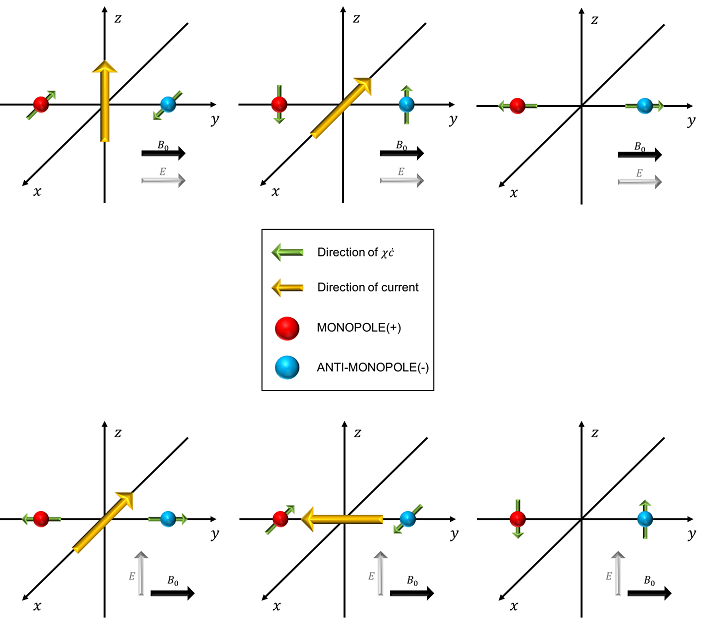}
\caption{Direction of the anomalous Hall current [Eqs. (\ref{AHE_BE_Berry_Electric_Field}) and (\ref{AHE_E_Berry_Electric_Field})] driven by the Berry electric field}
\end{figure}

\section{Conclusion}

The Berry electric field is a novel ingredient in a Weyl metal phase. When the distance between a pair of Weyl points changes as a function of time, the Berry electric field arises. In this situation both the Berry magnetic field and Berry electric field are governed by the Berry-Maxwell equation [Eq. (\ref{Berry_Maxwell_Equation})]. This Berry electric field should be introduced into the topologically modified Drude model Eq. (\ref{Effective_Velocity_Weyl_Metal}). As a result, we revealed the existence of anomalous Hall effects proportional to the Berry electric field, which should be distinguished from ``conventional" anomalous Hall currents given by the Berry magnetic field. Current directions of such anomalous Hall effects are classified systematically in all possible cases.

\section*{Acknowledgement}

This study was supported by the Ministry of Education, Science, and Technology (No. NRF-2015R1C1A1A01051629 and No. 2011-0030046) of the National Research Foundation of Korea (NRF) and by TJ Park Science Fellowship of the POSCO TJ Park Foundation. This work was also supported by the POSTECH Basic Science Research Institute Grant (2016). We would like to appreciate fruitful discussions in the APCTP Focus program ``Lecture Series on Beyond Landau Fermi Liquid and BCS Superconductivity near Quantum Criticality" (2016). KS appreciates fruitful discussions and collaborations with experimentalists of Heon-Jung Kim, Jeehoon Kim, and M. Sasaki.

\end{document}